\documentclass[pra,twocolumn]{revtex4-1}
\usepackage{graphics,epsfig ,color,graphicx} 

\begin{document}
\date{\today}
\title[Dynamics of laser threshold crossing]{
Some fundamental considerations on the dynamics of class B laser threshold crossing}
\author{G.P. Puccioni$^{1}$, T. Wang$^{2,3}$, and G.L. Lippi$^{2,3}$ \\}
\address{
$^1$ Istituto dei Sistemi Complessi, CNR, Via Madonna del Piano 10, I-50019 Sesto Fiorentino, Italy\\
$^2$ Institut Non Lin\'eaire de Nice, Universit\'e de Nice Sophia
Antipolis, France\\
$^3$ CNRS, UMR 7335\\
1361 Route des Lucioles, F-06560 Valbonne, France
}
\email{Gian-Luca.Lippi@inln.cnrs.fr}

\begin{abstract}
With the help of a simple rate equation model, we analyze the intrinsic dynamics of threshold crossing for Class B lasers.  A thorough discussion of the characteristics and the limitations of this very commonly employed model, which provides excellent qualitative predictions on the laser behaviour, is offered.  Approximate solutions for the population inversion and for the field intensity, up to the point where the latter reaches macroscopic levels, are found and discussed, together with the associated characteristic times.  Numerical verifications test the accuracy of these solutions and confirm their validity.  A discussion of the implications on threshold dynamics is presented, together with the motivation for focussing on this -- nowadays most common -- class of lasers.
\end{abstract}
\maketitle

\section{Introduction}
Lasers have been around for over half a century, and although initially considered as a pure technical curiosity (or {\it a solution looking for a problem}, as they were often referred to in the 1960's), they have become ubiquitous.  As a consequence, lasers are nowadays studied from different points of view, depending on the kind of application which is the ultimate goal of the approach.  This paper gives a contribution to the description of the dynamics of the threshold crossing of a wide class of lasers, the so-called Class B lasers~\cite{Tredicce1985}, whose properties, although well-known, are often viewed from a practical, rather than fundamental point of view.  Thus, in order to help understand the basics of these devices, we are going to highlight their temporal response to the threshold crossing with the help of the simplest possible model.

Threshold is normally viewed as a static property, and is used as a concept of principle, regardless of the way the laser makes the transition from the {\it off-state} (below threshold) to the {\it on-state} (above threshold).  In the simplified description based on the sole accounting of the stimulated photons, these two states correspond also to the lack of emission and to the presence of light emission, respectively.  Of course, the reality is more complex, since some (incoherent) light is emitted even below threshold, but the simplified picture is justified both in terms of emission strength -- orders of magnitude weaker below than above threshold --, in terms of coherence of the radiation (incoherent below threshold -- i.e., spontaneous emission --, highly coherent above threshold) and in terms of directionality of the radiation (i.e., the appearence of the {\it collimated ``pencil" beam} above threshold).  All these considerations hold very well for the macroscopic lasers developed in the first decades of laser fabrication and still apply quite well to most current lasers, including semiconductor ones, in spite of their very small size~\cite{note1}. 

Before delving into the discussion, however, it is important to remark on  the choice of the Class B laser for this discussion.  The latter represent the ensemble of lasers whose description is based on two main physical variables:  the electromagnetic (e.m.) field intensity and the population inversion.  The simplest device of this kind will therefore be a single longitudinal and transverse mode laser, possessing a ring cavity with unidirectional emission (i.e., the so-called {\it unidirectional ring cavity}).  The latter can be obtained by inserting a non-reciprocal element (e.g., a Faraday isolator~\cite{Saleh2007}) inside a ring cavity.  

Compared to the Class A laser, physically described by the sole e.m. field intensity (e.g., He-Ne lasers and gas lasers in general, dye lasers $\ldots$), the Class B device possesses a dynamics which is determined by the interplay of the physics of the e.m. field and that of the material's response (restricted to the difference in population between the upper and lower state participating in the lasing transition)~\cite{note2}.  To this class belong notably semiconductor lasers, solid state lasers and some molecular lasers (e.g., the CO$_{\rm 2}$, which finds applications both in medicine -- surgery or heat treatment -- and in technology -- cutting and drilling). 

While the more complex interaction presents a fundamental interest for the study of the dynamics of these lasers, at the same time class B lasers represent the single most important source of coherent emission for technological applications, accounting for well over 90\% of the World laser sales~\cite{SPIE2010} (2010 data).  This practical interest is closely related to their physical properties since to the Class B belong lasers whose material time constant (i.e., the timescale on which the population inversion reacts) is slower than the e.m. field's.  This implies that the spontaneous losses, due to relaxation of the upper state without contributing a photon to the stimulated component, are strongly reduced; in other words, these devices intrinsically possess much lower losses and offer an efficiency which can be orders of magnitude larger than that of Class A lasers.   Thanks to this intrinsic and unmistakeable advantage, technological solutions have been developed over the years to offer class B lasers emitting virtually on all wavelengths from mid-IR to near-UV.  Thus, in addition to the fundamental interest in studying the more complex dynamics, we find the practical motivation of understanding threshold crossing as it occurs in those lasers which are used in almost all of everyday's applications.

\section{Model properties}

There exist numerous derivations of the basic model for a laser, and the Class B model~\cite{Arecchi1965} can be obtained from the Maxwell-Bloch\cite{note3} model performing the adiabatic elimination of the atomic polarization (cf. e.g.,~\cite{Narducci1988}).  However, it is easy to show that they all reduce to the so-called rate equations, which can be written directly from physical considerations~\cite{Siegman1986}:
\begin{eqnarray}
\label{rateeqn}
\frac{d n}{ d t} & = & - \kappa n + G n N \, , \\
\label{rateeqN}
\frac{d N}{d t} & = & R - \gamma N - G n N \, ,
\end{eqnarray}
where $n$ represents the photon number, $N$ the difference between the number of {\it atoms}~\cite{note4} in the upper and in the lower level of the lasing transition, $\kappa$  represents the losses for the photon number, $G$ the coupling constant between photons and {\it atomic} excitation, $R$ stands for the pump rate (i.e., amount of energy supplied per unit time to the laser) and $\gamma$  the spontaneous relaxation rate of the population from the upper state.  In writing equations~(\ref{rateeqn}-\ref{rateeqN}) we have made the implicit assumption that we are considering a perfect four-level laser, with infinitely fast relaxation from the lower state towards a separate, fundamental state~\cite{Siegman1986}.  This assumption does not qualitatively change the results of our analysis and is justified by its simplicity.  Quantitative changes are discussed by several authors (cf. e.g., Siegman's book~\cite{Siegman1986}).

\begin{table}
\caption{
Transformations between the two forms of model (equations~(\ref{rateeqn}--\ref{rateeqN}) vs. equations~(\ref{rateeqI}--\ref{rateeqD})).
}
\label{transformations}
\begin{tabular}{||c  c  c ||}\hline\hline
Direct physical model & & Normalized model\\\hline\hline
$n$ & $\leftrightarrow$ & $\frac{\gamma}{2 G} I$ \\ \hline
$N$ & $\leftrightarrow$ & $\frac{K}{G} D$ \\ \hline
$R$ & $\leftrightarrow$ & $\frac{\gamma K}{G} P$ \\ \hline
$\kappa$ & $\leftrightarrow$ & $K$\\ \hline\hline
\end{tabular}
\end{table}

With the help of the transformations detailed in Table~\ref{transformations} it is possible to recast the rate equations into a normalized form, more suitable for our discussion, as follows:
\begin{eqnarray}
\label{rateeqI}
\frac{d I}{ d t} & = & - K (1 - D) I \, , \\
\label{rateeqD}
\frac{d D}{d t} & = & - \gamma [(1+I) D - P] \, ,
\end{eqnarray}
where $I$ stands for the e.m. field intensity, $D$ for the population inversion, $K$ for the intensity losses, and $P$ represents the pump (i.e., energy supplied to the laser).

The recast version of the rate equations immediately highlights the existence of the two time scales:  the e.m. field intensity evolves over a timescale $\tau_I \sim \frac{1}{K}$ (equation~(\ref{rateeqI})), while the population inversion's timescale is $\tau_D \sim \frac{1}{\gamma}$ (equation~(\ref{rateeqD})).  This directly illustrates the physical characteristics which identify class B lasers.  Typical values of the relaxation constants are offered in Table~\ref{relax} for some selected sample devices.

\begin{table}
\caption{
Some typical relaxation constants valid for some selected, sample class B lasers.
}
\label{relax}
\begin{tabular}{|| c | c | c ||}\hline\hline
Type of laser & $K (s^{-1})$ & $ \gamma (s^{-1})$ \\ \hline\hline 
CO$_ {\rm 2}$ & $10^7$ & $10^4$ \\ \hline
Nd:YAG & $10^8$ & $10^4$\\ \hline
Semiconductor & $10^{11}$ & $10^9$ \\ \hline\hline
\end{tabular}
\end{table}

These equations straightforwardly possess the following double set of steady-state solutions~\cite{Narducci1988} (the overstrike denoting the steady state):
\begin{eqnarray}
\label{ss}
\left(
\begin{array}{c}
\overline{I} = 0\\
\overline{D} = P\\
\end{array}
\right) 
& \quad , \quad & 
\left(
\begin{array}{c}
\overline{I} = P-1\\
\overline{D} = 1\\
\end{array}
\right) 
\end{eqnarray}
where the threshold value in the normalized form of this model is $P_{th} = 1$ (i.e., $\overline{I} = 0$).

A linear stability analysis of the above-threshold solution~\cite{Narducci1988,Mandel1997} immediately provides stable solutions (for $P \ge 1$) with eigenvalues of the form:
\begin{eqnarray}
\label{eigenv}
\lambda & = & \frac{1}{2} \left[ - \gamma P \pm \sqrt{\gamma^2 P^2 - 4 \gamma K (P-1)} \right] \, ,
\end{eqnarray}
where the square root takes imaginary values as soon as 
\begin{eqnarray}
P & \gtrsim & 1 + \frac{1}{4} \frac{\gamma}{K} \, .
\end{eqnarray}
Given that $\gamma \ll K$ for all Class B lasers, the eigenvalues, equation~(\ref{eigenv}), are (almost) always complex above threshold, and represent a (damped) oscillation with angular frequency 
\begin{eqnarray}
\label{omegar}
\omega \approx \sqrt{\gamma K (P-1)} \, ,
\end{eqnarray}
where obtaining this approximate expression we have neglected the term $\gamma^2 P^2$, very small compared to $4 \gamma K (P-1)$ for all practical values of $P$.

Notice that the model we are studying is entirely deterministic and does not take into account in any way the presence of spontaneous emission.  In other words, the model accounts only for the stimulated fraction (i.e., perfectly coherent) of the emitted photons and entirely ignores the spontaneous one (negligible once threshold is attained -- cf. section~\ref{num} for numerical estimates).  An improvement on this model is represented by a set of rate equations where the average contribution of the spontaneous emission is accounted for in the field intensity equation (cf. e.g.,~\cite{Coldren2012}).  This addition, however, does not qualitatively alter the physical description, while it adds a good degree of mathematical complexity (cf., e.g., the imperfect bifurcation problem in laser physics~\cite{Erneux1986}).  We therefore use the simpler model (equations~(\ref{rateeqI}--\ref{rateeqD})) paying close attention to the interpretation of its predictions and to the use of the boundary conditions (cf. discussion of the value of $\tilde{I}$ in section~\ref{predictions}).

\section{Dynamical threshold crossing}\label{predictions}

When considering the transition from below to above threshold we explicitely deal with the condition $P(t=0^-) < 1$ and therefore $I(t=0) = 0$ and $D(t=0) \equiv D_0 = P(t=0^-)$, according to the first set of steady-state values, equation~(\ref{ss}).  Assuming a Heaviside function shape for the pump ($P(t<0) < 1, P(t>0) > 1$), the model reduces to
\begin{eqnarray}
\frac{d I}{d t} & = & 0 \, , \\
\frac{d D}{d t} & = & -\gamma (D - P) \, ,
\end{eqnarray}
with the initial conditions specified above.  Thus, the e.m. field intensity remains constant, at zero, while the population inversion starts growing exponentially according to
\begin{eqnarray}
\label{initialD}
D(t) & = & (P - D_0) ( 1 - e^{- \gamma t} ) + D_0 \, ,
\end{eqnarray}
which holds until the instant $\tilde{t}$ at which the population inversion reaches the other solution, equations~(\ref{ss}):  $D(\tilde{t}) = 1$.  Starting from this instant, the first of the model equations (\ref{rateeqI}) acquires a positive r.h.s. and the e.m. field intensity starts to grow.  

The value of $\tilde{t}$ can be easily obtained from equation~(\ref{initialD}):
\begin{eqnarray}
D(\tilde{t}) & = & 1 = (P - D_0) ( 1 - e^{- \gamma \tilde{t}} ) + D_0 \, , \\ 
\label{t-tildedef}
\tilde{t} & = & - \frac{1}{\gamma} \log \left[ \frac{P-1}{P-D_0} \right] \, ,
\end{eqnarray}
where we are assured that $\tilde{t} > 0$ by the fact that $P-1 < P - D_0$.  Starting from this instant, the full model, equations~(\ref{rateeqI}-\ref{rateeqD}), must be used in its nonlinear form and no closed solution exists for the time evolution of the physical variables.  However, if we concentrate on the initial phases of the e.m. field intensity growth, we can gather some information on the timescale over which the laser turns on.

Assuming that the laser intensity $I$ is small in its initial phases ($I \ll 1$), we can suppose its influence on the evolution of $D$ to be negligible (since $(1+I) \approx 1$ in the r.h.s. of equation~(\ref{rateeqD})) and obtain an approximate form for $D(t > \tilde{t})$:
\begin{eqnarray}
D(t) & = & D(\tilde{t}) + D(\delta t) \, , \\
\label{Dabovethr}
& = & 1 + (P-1) \left( 1 - e^{- \gamma \, \delta t} \right) \, , \qquad \delta t \equiv t- \tilde{t} \\
& \approx & 1 + (P-1) \left[1 - (1 - \gamma \, \delta t) \right] \, , \\
\label{approxD}
& = & 1 + (P-1) \gamma \, \delta t \, ,
\end{eqnarray}
which holds as long as $\delta t \ll \frac{1}{\gamma}$, a condition which is very well satisfied in practice (and which can be easily checked {\it a posteriori} -- cf. section~\ref{num}).

We can now use this approximate solution to get an approximate solution for the initial phases of the e.m. field intensity growth by replacing $D(t)$ from equation~(\ref{approxD}) into the rate equation for $I$ (equation~(\ref{rateeqI})), which can be easily recast as:
\begin{eqnarray}
\frac{d (\log I)}{d t} & = & \gamma K (P-1) \delta t \, .
\end{eqnarray}

Direct integration provides
\begin{eqnarray}
\label{formalintegral}
\int_{\tilde{t}}^t d(\log I) & = & \gamma K (P-1) \int_{\tilde{t}}^t  (t^{\prime} - \tilde{t}) d t^{\prime} \\
& = & \frac{1}{2} \gamma K (P-1) (t - \tilde{t})^2 \, ,
\end{eqnarray}
which only holds until a time $t_M$, to be determined.  The l.h.s. of equation~(\ref{formalintegral}) provides $\log I(t)$ up to a constant ($\log I(\tilde{t})$) which corresponds to a mathematical divergence, since $I(\tilde{t}) = 0$.  Besides being unphysical, this is an artefact of the model, which considers only the deterministic evolution of the coherent fraction of the e.m. field:  spontaneous emission is not included in this description.  A self-consistent solution can only be obtained by including the spontaneous photons in the description, but the complexity of the model increases considerably;  the average properties of the lasing transition, however, are still correctly given by the set of rate equations~(\ref{rateeqI}-\ref{rateeqD}). 
Thus, we can use the correct physical condition (i.e., the average value of the spontaneous emission in the lasing mode at threshold) to estimate the value of $I(\tilde{t})$, thus avoiding the unphysical divergence.  Indicating with $I_0$ this value (i.e., the value of $I$ at $t = \tilde{t}$), we obtain the approximate expression for the e.m. field intensity growth:
\begin{eqnarray}
\label{approxI}
I(t) & = & I_0 e^{\frac{1}{2} \gamma K (P-1) (t - \tilde{t})^2} \, ,
\end{eqnarray}
which already provides us with a wealth of (deterministic) information about the initial phases of the growth of the laser intensity:
\begin{itemize}\itemsep0cm
\item[$\bullet$] the growth is exponential but with a quadratic time dependence -- since the solution~(\ref{approxI}) holds only at short times, the quadratic growth in time indicates a slower initial growth than what would result from a linear time dependence;
\item[$\bullet$] the speed at which the laser intensity grows depends on the distance of the pump from threshold ($P-1$) -- the larger the pump, the faster the growth;
\item[$\bullet$] the time constant for the intensity growth is not proportional to $K^{-1}$, as one would, mistakenly but intuitively, expect from the timescale evolution of the intensity (cf. equation~(\ref{rateeqI})), but rather to the geometric mean of the two time constants $(\gamma K)^{-1/2}$;  
\item[$\bullet$] the actual time constant for the exponential growth is $\tau_{exp} = \sqrt{\frac{2}{\gamma K (P-1)}}$, i.e., the distance of the pump to threshold induces a hyperbolic lengthening of the time scale, with the usual divergence, typical of critical slowing down~\cite{Haken1983}, taking place as $P \rightarrow 1$.
\end{itemize}
While intuitively unexpected, the dependence of the timescale on the product of the two relaxation constants for the physical variables is logical.  Indeed, it is not sufficient for the e.m. field intensity to grow at a rate $K^{-1}$ since the population inversion must have the time to increase as well in order for the photon number to develop.
Notice that the relaxation oscillations, equation~(\ref{omegar}), appearing around the above-threshold solution (thus, far beyond the intensity ranges we are considering here) have the same parameter dependence as the time delay $\Delta t \equiv t - \tilde{t}$, apart from a numerical coefficient.  This point is significant since it shows how the time constants appearing in all parts of the transient evolution are closely related to each other by the intrinsic physical interplay between the laser variables.

The limits of validity of the solution we have obtained for the transient can be easily established in the following way.  The transient dynamics of class B lasers is characterized by a delay in the laser intensity growth, accompanied by an overshoot of its value beyond its asymptotic state with damped oscillations~\cite{Coldren2012}.  A dynamical analysis can be performed, looking, among others indicators, at the shape of the trajectory in phase  space~\cite{Lippi2000}.  Strong deviations for the growth of the population inversion from the approximate solution, equation~(\ref{approxD}), are expected, and numerically found, when the laser intensity exceeds its asymptotic value $\overline{I} = P-1$.  Thus, we can set the limit of validity at a fraction of this value $a \overline{I}$ ($a<1$, arbitrary) to determine the maximum time value $t_M$ for which the approximate solution for $I(t)$ (equation~(\ref{approxI})) holds:
\begin{eqnarray}
a (P-1) & = & I_0 e^{\frac{1}{2} \gamma K (P-1) (t_M - \tilde{t})^2} \, ,
\end{eqnarray}
which immediately gives an estimate for $t_M$:
\begin{eqnarray}
\label{exprtM}
t_M & = & \tilde{t} + \sqrt{\frac{2}{\gamma K (P-1)} \log \left( \frac{a (P-1)}{I_0} \right) } \, .
\end{eqnarray}
Since we are trying to estimate the time necessary to attain a fraction $a$ of the steady state value for the laser intensity, it does not make sense to consider the limit $P \rightarrow 1$, thus the potential divergences present in the expression on the r.h.s. of equation~(\ref{exprtM}) lie outside the realm of the interesting physical parameter ranges.

\section{Numerical verifications}\label{num}

\begin{figure}[ht!]
\includegraphics[width=0.9		
\linewidth,clip=true]{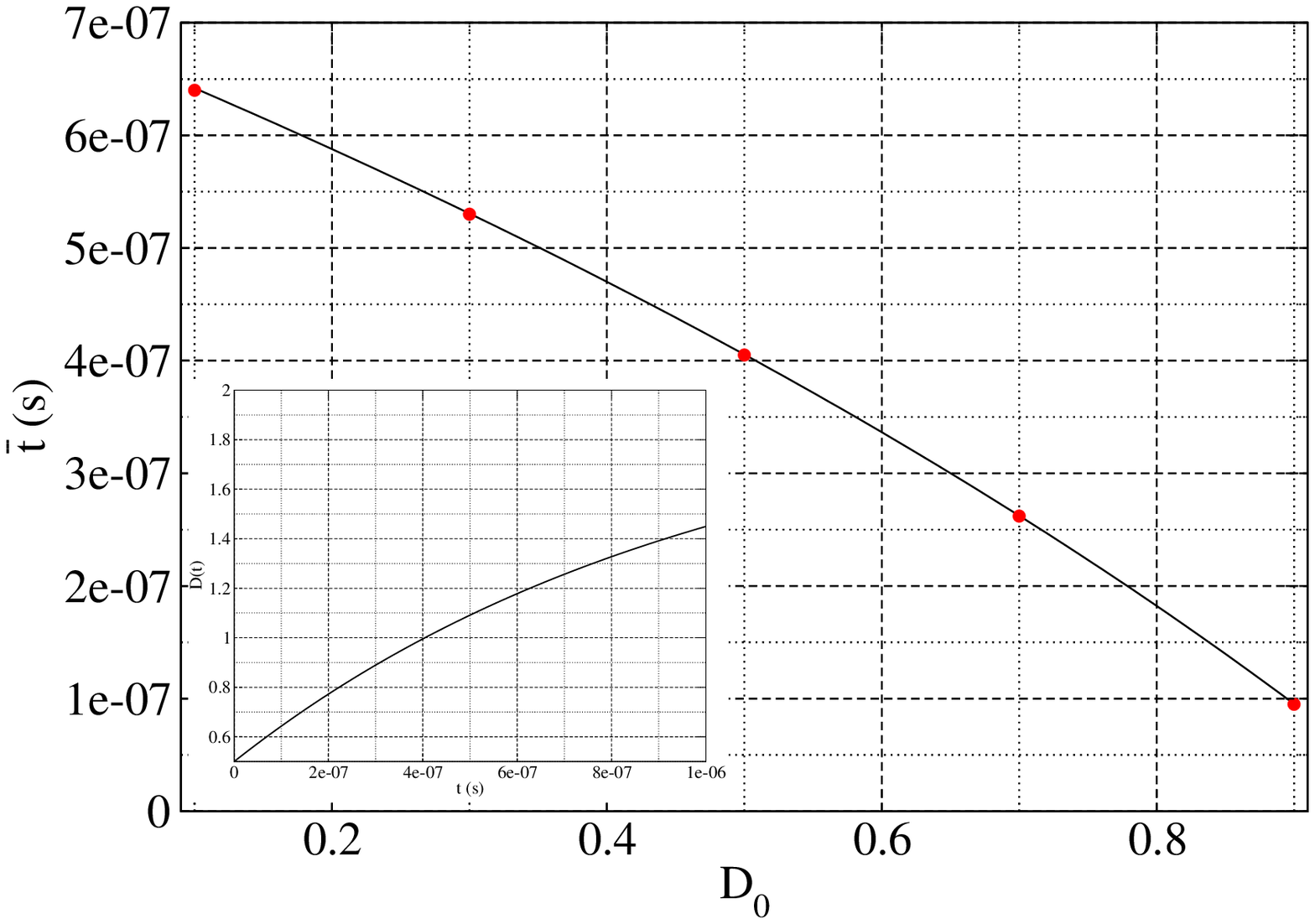}
\caption{
Comparison between the value of the time value at which $D=1$ occurs obtained from equation~(\ref{t-tildedef}) -- solid line -- and the equivalent values obtained from the numerical integration of the model (equations~(\ref{rateeqI},\ref{rateeqD})) -- dots.  For this and the following figures the following parameter values are used:  $\gamma = 1 \times 10^6 s^{-1}$, $K = 1 \times 10^8 s^{-1}$, $P = 2$.  The inset shows the temporal evolution of the population inversion computed from equation~(\ref{initialD}) for a time equal to $\frac{1}{\gamma}$; the asymptotic value is at $P=2$.
} 
\label{t-tilde} 
\end{figure}

A verification of the approximate solutions is easily obtained by comparing the analytical predictions to the numerical values resulting from the integration of the model, equations~(\ref{rateeqI},\ref{rateeqD}), obtained with a first-order Euler scheme programmed in GNU Octave. 
The temporal evolution of the population inversion (cf. inset of Fig.~\ref{t-tilde}) displays a growth corresponding to that of a saturating exponential, as predicted by equation~(\ref{initialD}).  The crossing time $\tilde{t}$ ($\tilde{t}$:  $D(\tilde{t}) = \overline{D} = 1$) can be easily found from this trajectory (and more precisely from the numerical file).  We also remark that, as implicit in the previous discussion, the population inversion $D$ grows beyond its asymptotic value in the process of laser threshold crossing (cf. discussion in section~\ref{discussion}).

\begin{figure}[ht!]
\includegraphics[width=0.9		\linewidth,clip=true]{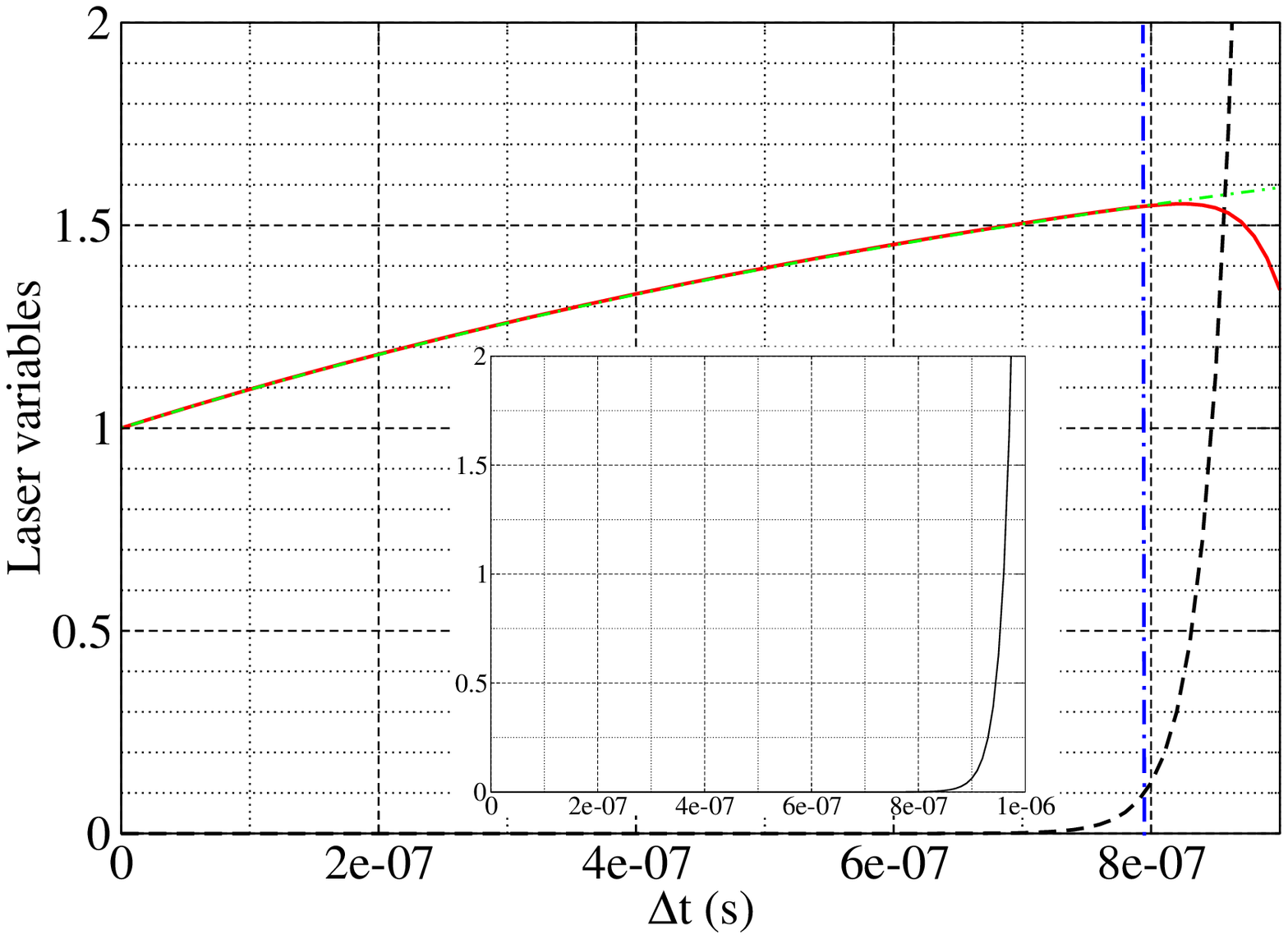}
\caption{
Time evolution of the laser variables as a function of $\Delta t = t - \tilde{t}$:  Laser intensity (dashed line -- black online), population inversion (continuous line -- red online), approximate solution (equation~(\ref{Dabovethr})) for the population inversion (dash-double dotted line -- green online).  The vertical dash-dotted line (blue online) marks the value of $t_M$:  the approximate and the exact solution are still an excellent match up until this time value.  Inset:  shape of the field intensity growth according to the approximate solution, equation~(\ref{approxI}).  The agreement is qualitative but shows that the shape of the curve is well reproduced -- notice that the value of $\gamma K$ has been replaced by $\gamma K/2$ for this graph (cf. text for details).  The initial value used for the the field intensity (representing the average spontaneous emission) is $I_0 = 1 \times 10^{-10}$.
} 
\label{time-evolution} 
\end{figure}

Fig.~\ref{t-tilde}  shows the comparison between the time $\tilde{t}$ necessary for the population inversion to reach its above-threshold steady state value (continuous line) as predicted from equation~(\ref{t-tildedef}),  and the numerical time obtained from the temporal trajectory (dots).  Not unexpectedly, since the approximation used in the derivation of equation~(\ref{t-tildedef}) is very well verified, the agreement is excellent.  

Once threshold is crossed, the numerical integration has to start from a good estimate of the ``initial" value of the laser field intensity.  The deterministic rate equation model does not account for the background {\it noisy} dynamics which holds the intensity value constant (in average) around the value of the spontaneous emission.  If one starts the integration of equations~(\ref{rateeqI}-\ref{rateeqD}) with a deterministic initial condition (e.g., $D_0 = 0.5$, as in one of the simulations run for Fig.~\ref{t-tilde}), during the whole transient where $D(t) < \overline{D} = 1$, the laser intensity decays away to ever smaller numbers.  This is an artefact of the model and should not be mistaken for a physical effect.  Continuing the simulation from unphysically low values of the laser intensity (e.g., much lower than the average spontaneous emission level) would artificially increase the latency time needed to reach macroscopic intensity values, and thus affect the maximum values reached by the population inversion, and, as a consequence, by the laser intensity at its peak (not discussed here -- the full time evolution can be seen, for instance, in Ref.~\cite{Lippi2000}).  Thus, it is crucial to consider a reasonable estimate of the average spontaneous emission.  Traditionally, the following physical considerations have been employed to conceptually define threshold:  for the stimulated emission to overcome the spontaneous emission and concentrate on the lasing mode all (or most) of the energy, the number of photons in the lasing mode has to equal the number $N$ of modes available for the spontaneous emission (i.e., the number of electromagnetic cavity modes).  In other words, while in average the number of spontaneous photons is $\langle n_{s,j} \rangle = 1$ for each mode ($j = 1 \ldots N$), the (average) number of stimulated photons in mode $p$ must be $\langle n_{st,p} \rangle = N$ for lasing action to occur~\cite{note5}.  Thus, if we consider a laser whose cavity possesses $N$ modes, its relative average spontaneous intensity {\it at threshold} will be $\frac{\langle I_{sp} \rangle}{\overline{I}} = \frac{1}{N}$.  Without entering into details, macroscopic lasers have values of $10^7 < N < 10^{12}$ (and even beyond); small semiconductor lasers are characterised by $N \approx 10^5$, while smaller cavities exit the realm of small-sized lasers to approach the nanoscale.  A more detailed discussion, supported by stochastic calculations, can be found in~\cite{Wang2015,Puccioni2015}.  Here we use values $10^5 \le N \le 10^{10}$, specified in the figures as appropriate.

The evolution of the population inversion following $\tilde{t}$ is displayed in Fig.~\ref{time-evolution}.  The continuous lines (red online) shows the population inversion numerically integrated from the model, equations~(\ref{rateeqI}-\ref{rateeqD}), while the dashed line (green online) represents the predictions of the approximate expression, equation~(\ref{initialD}).  The graph convincingly shows that the analytical approximation holds well beyond $\tilde{t}$, even once the laser intensity $I$ starts growing away from $0$.  Indeed, the two curves are superposed for times exceeding $t = 8 \times 10^{-7} s$ (for the parameter values of the figure), and remain very close until $I \approx \frac{\overline{I}}{2}$ ($\frac{\overline{I}}{2}$ = 1 for the chosen parameters).

\begin{figure}[ht!]
\includegraphics[width=0.9		\linewidth,clip=true]{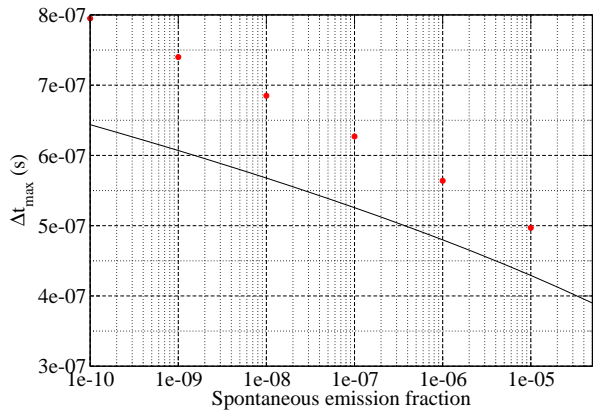}
\caption{
Comparison between the value of the time value for which $I(t_M)=a \overline{I}$ (plotted as $\Delta t_{max} = t_M - \tilde{t}$) as a function of the value of the spontaneous emission ($I_0$).  Solid line:  approximate expression -- equation~(\ref{exprtM}) --; dots:  numerical integration of the model (equations~(\ref{rateeqI},\ref{rateeqD})).  Cf. text for details.  The discrimination level for the intensity has been set at $a = 0.1$ (i.e., 10\% of the asymptotic value $\overline{I}$, according to equation~(\ref{ss}) -- right hand group of solutions).
} 
\label{deltatmax} 
\end{figure}

The analytical predictions of section~\ref{predictions} have provided also an estimate of the maximum time value for which the approximate analysis holds.  Fig.~\ref{deltatmax} shows a comparison between the estimated time, as a function of the spontaneous emission fraction (i.e., $\langle I_{sp,p} \rangle / \overline{I}$).  The agreement here is somewhat less good than the one previously obtained, due to the fact that we have retained only the linear term (first-order correction) in the expressions for the population inversion, equation~(\ref{approxD}), to obtain an approximate behaviour for the initial phases of the intensity growth, as reproduced by equation~(\ref{approxI}).  It is from this latter equation which we have estimated the maximum time, equation~(\ref{exprtM}), represented as a time difference $\Delta t_{max} \equiv (t_M -\tilde{t})$ in Fig.~\ref{deltatmax}.  Notice, however, that the order of magnitude is correctly obtained and that the largest error is of the order of 20\%:  it occurs, not surpisingly, for the lower values of the spontaneous emission, which lead to longer values of $t_M$.

We also remark that the threshold set for determining the value of $t_M$ ($a = 0.1$) falls well within the range of validity for the approximate expression of the population inversion given by equation~(\ref{approxD}):  the time $t_M$ is marked in Fig.~\ref{time-evolution} by the vertical dot-dashed line (blue online) -- at this instant, the numerical and the analytical expression for $D(t_M)$ coincide.

Finally, we look at the shape of the initial growth of the field intensity, as displayed in the inset of Fig.~\ref{time-evolution} for comparison with the dashed line (black online) in the same figure.  The overall shape is quite well reproduced, even surprisingly well for an approximate solution with a growth rate as large as that of a quadratic exponential, but for the value of the time constant $\tau_{exp}$ which, for the sake of graphical comparison, has been doubled.  Given the rather crude approximations used to obtain the shape of the growing intensity, equation~(\ref{approxI}), then used for estimating the time $t_M$, the qualitative agreement is quite satisfactory.  As a last remark, the value of the delay time $\tau_{exp}$ used in the numerical comparison, larger than the one coming from the analytical estimate, brings its value a bit closer to the actual response time (Fig.~\ref{deltatmax}) and to the relaxation oscillation period, estimated from the linear stability analysis, equation~(\ref{omegar}).

\section{Discussion and conclusions}\label{discussion}

The usual picture of laser threshold is based on a static representation, where the field becomes coherent as soon as the pump rate exceeds its threshold value.  This picture rests on the validity of the assumption of an infinitely large system (i.e., the thermodynamical limit~\cite{Dohm1972} for the laser), which is very well satisfied by a large class of existing devices:  in practice all solid state lasers (even microdisks) and all traditional gas lasers, high power lasers, etc.~\cite{note6}.
Refinements become necessary when studying different laser classes.  Recent work has shown that the well-established characterization of coherence properly holds only for Class A devices~\cite{Wang2015} and in these systems a dynamical perspective in the crossing of threshold does not reserve particular surprises.  Due to the restricted (one-dimensional) phase space, in such systems the evolution of the field intensity takes a monotonic form and the only interesting aspects cover the delay time associated with the threshold crossing due to the time-dependence of the control parameter~\cite{Mandel1997,Scharpf1987}.

The larger phase space associated with the physics of the Class B laser, instead, renders the dynamics nontrivial, since it allows for a non-monotonic evolution of the laser intensity, in addition to the appearance of an intrinsic time delay, superposed to the one induced by the bifurcation~\cite{Mandel1997,Tredicce2004}.  This intrinsic delay stems from the fact that: 1. the field intensity cannot grow until the population inversion has reached its threshold value, and 2. a sufficient amount of inversion is needed to allow for the growth of the photon number.  This introduces a causal element which requires the population to grow by a sufficient amount for the field intensity to approach its above-threshold value; it is also the cause for the appearance of a timescale proportional to the geometric mean of the two relaxation constants ($\gamma$ and $K$).  In the absence of this causal component, one should have expected the field intensity to grow at a rate controlled by $K$, once the population inversion has reached its threshold value.

Summarizing the results of this paper, we have obtained approximate expressions for the times at which the population inversion reaches its threshold and the field intensity attains macroscopic values, together with approximate solutions for both variables within the time intervals just defined.  The agreement is quite satisfactory in all cases (and even excellent in some), in spite of the extreme simplicity of the analysis.  These considerations allow for a deeper insight into the threshold crossing properties of Class B lasers.

\section*{Acknowledgments}
GLL is grateful to all the students that have taken the graduate course in Laser Dynamics for the stimulating discussions and questions and to all the collegues with whom, in the past decades, he has had the opportunity to discuss issues related to laser transients.
\ \\[3mm]

\end{document}